\documentclass[letterpaper, 10 pt, conference]{ieeeconf}  % Comment this line out
\IEEEoverridecommandlockouts                              % This command is only
\usepackage{graphicx}
\usepackage{pslatex}
\usepackage{indentfirst}
\usepackage{amssymb}
\usepackage{lipsum}
\usepackage{mathtools}
\usepackage{cuted}

\title{{\Large \bf Degrees of Freedom of Spatial Self-Interference Suppression for In-Band Full-Duplex with Inter-node Interference}}
 
\author{{\large \bf Yujun Chen and Ashutosh Sabharwal} \\
  Department of Electrical and Computer Engineering, Rice University, Houston, TX, 77005 \\
  Email:\{yc67, ashu\}@rice.edu}
  
\begin{document}

\maketitle
\thispagestyle{empty}
\pagestyle{empty}

%%%%%%%%%%%%%%%%%%%%%%%%%%%%%%%%%%%%%%%%%%%%%%%%%%%%%%%%%%%%%%%%%%%%%%%%%%%%%%%%
\begin{abstract}

We study a three-node network with a full-duplex base-station communicating with one uplink and one downlink half-duplex node. In this network, both self-interference and inter-node interference are present. We use an antenna-theory-based channel model to study the spatial degrees of freedom of such network, and study if and how much does spatial isolation outperforms time-division (i.e. half-duplex) counterpart. Using degrees of freedom analysis, we show that spatial isolation outperforms time division, unless the angular spread of the objects that scatters to the intended users is overlapped by the spread of objects that back scatters to the receivers.

\end{abstract}

%%%%%%%%%%%%%%%%%%%%%%%%%%%%%%%%%%%%%%%%%%%%%%%%%%%%%%%%%%%%%%%%%%%%%%%%%%%%%%%%
\section{INTRODUCTION}

\begin{figure}[h]
\centering
\includegraphics[scale=0.3]{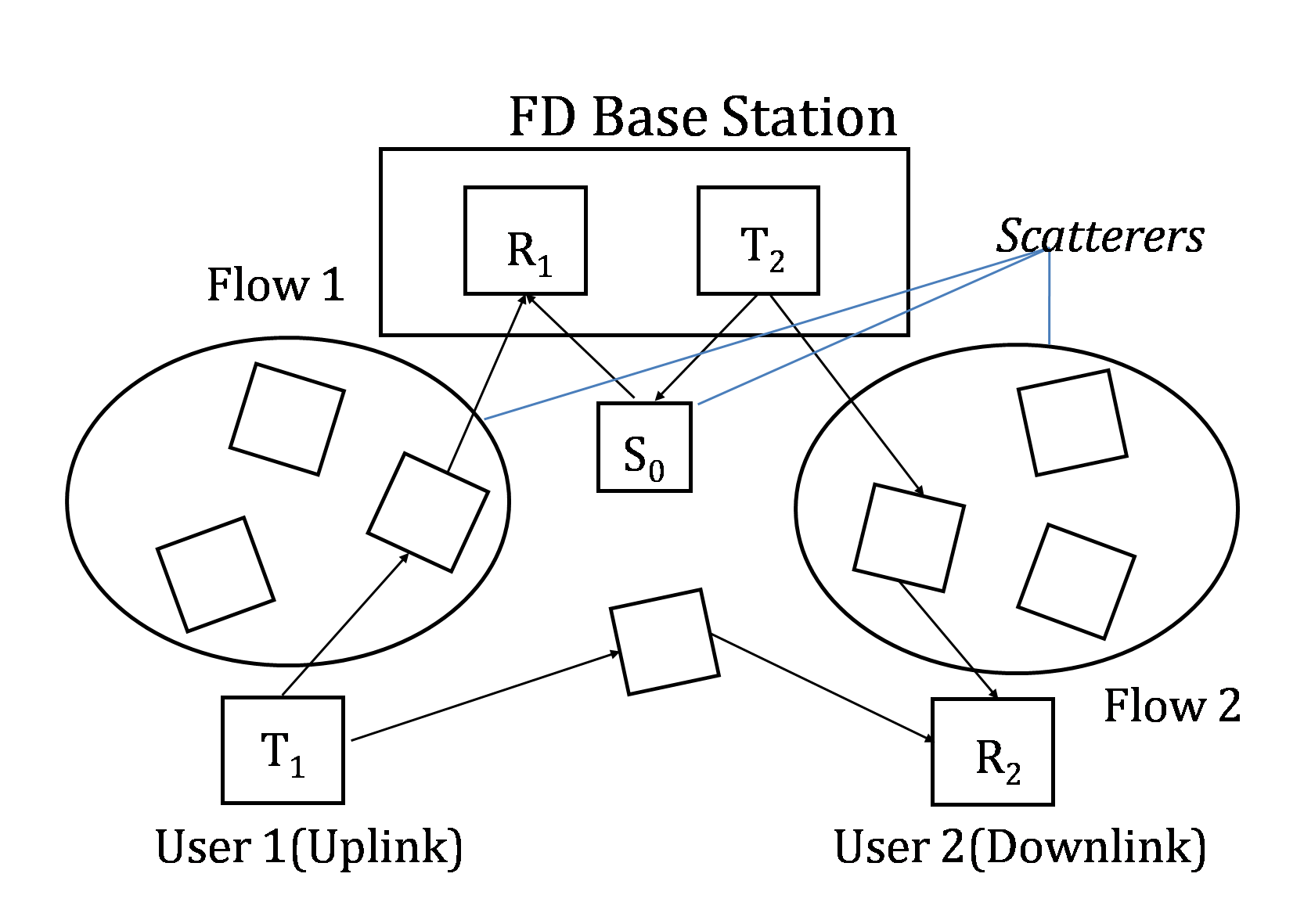}
\caption{Clustered scattering network \label{fig:network}}
\end{figure}

In networks where nodes can use in-band full-duplex with half-duplex users, see for example Figure~\ref{fig:network}, the network faces both self-interference and inter-node interference. The self-interference is caused by the base-station's own transmissions, and the inter-node interference at the downlink node is caused by uplink transmissions of User~1. The two forms of interference may reduce the overall capacity gain of full-duplex capability at the base-station. In this paper, we derive a signal-space degrees-of-freedom region for the three-node network shown in Figure~\ref{fig:network}. 

   %\begin{figure}[thpb]
   %   \centering
   %   \framebox{\parbox{3in}{We suggest that you use a text box to insert a graphic (which is ideally a 300 dpi TIFF or EPS file, with all fonts embedded) because, in an document, this method is somewhat more stable than directly inserting a picture.
%}}
 %     \includegraphics[scale=0.3]{system_ver3}
  %    \caption{Clustered scattering}
   %   \label{network}
  % \end{figure}

    To solve the problem of self-interference and inter-node interference, it is possible to use analog/digital cancellation in combination with spatially suppression; spatially suppression of self-interference can be obtained by increasing the path-loss between transmit and receive antennas, either by fixed methods (e.g.\ separation or isolation) or adaptive methods (e.g.\ beamforming). Cancellation based methods use prior knowledge of the transmit signal to subtract an estimate of the self-interference at the receiver side ~\cite{c1}. Spatial suppression reduces the self-interference by spatially orthogonalizing the self-interference and the signal-of-interest~\cite{c2}. Previous studies have shown that spatial suppression is an effective technique~\cite{c3}, and thus we study degrees-of-freedom due to spatial suppression in this paper. 
    
    This work is an extension of our  prior work~\cite{c4}, where we considered Figure~\ref{fig:network} without inter-node interference. In this paper, we consider both self- and inter-node interference, and understandably the general problem is more challenging. Through analyzing the half-duplex and full-duplex with self- and inter-node interference cases, we conclude that both self- and inter-node interference hinder us from achieving the maximum full-duplex gain.
 Both interference can be reduced by increasing the array sizes of receivers and transmitters.

\section{System Model}

We use the antenna-theory-based channel model developed by Poon, Broderson, and Tse~\cite{c5} to analyze this full-duplex system, and we call it the PBT model. The PBT channel model considers a wireless communication link between a transmitter with a unipolarized continuous linear array of length $2L_T$ and a receiver with a similar array of length $2L_R$. There are two key domains: the array domain, which describes the current distribution on the arrays, and the wave vector domain, which describes radiated and received field patterns. The authors of ~\cite{c5} focus on the union of the clusters of departure angles from the transmit array, denoted as $\Theta_T$, and the union of the clusters of arrival angles to the receive array, $\Theta_R$. Because a linear array aligned to the z-axis can only resolve the z-component, the intervals of interest are $\Psi_T = \{cos\theta: \theta \in \Theta_T\}$ and $\Psi_R = \{cos\theta: \theta \in \Theta_R\}$.It is shown from the first principles of Maxwell's equations that an array of length $2L_T$ has a resolution of $1/(2L_T)$ over the interval $\Psi_T$, so that the dimension of the transmit signal space of radiated field pattern is $2L_T|\Psi_T|$. Similarly the receive signal space is $2L_R|\Psi_R|$, so that the degrees of freedom of the communication link is $d_{P2P}=min\{2L_T|\Psi_T|,2L_R|\Psi_R|\}$.

Now we extend this PBT channel model to the three-node full-duplex topology of Figure 1. Let Flow 1 denote the uplink flow from User 1 to the base station, and $T_1$ be User 1's transmitter and $R_1$ be the base station's receiver. Similarly, let Flow 2 denote the downlink flow from the base station to User 2, and $T_2$ be the base station's transmitter, and $R_2$ be User 2's receiver.

\begin{figure}[h]
\centering
\includegraphics[scale=0.5]{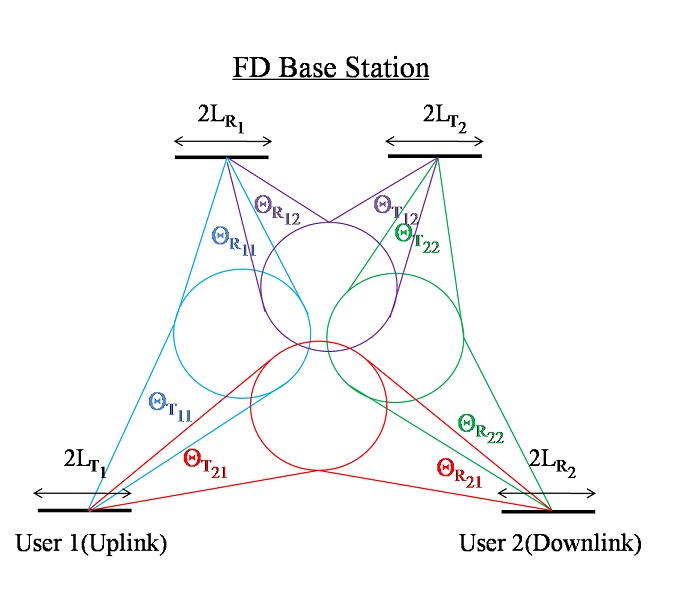}
\caption{Clustered scattering}
\end{figure}

We consider continuous linear arrays of infinitely many infinitesimally small unipolarized antenna elements. Each of the two transmitters $T_j,j=1,2$, is equipped with a linear array of length $2L_{T_j}$, and each receiver $R_i, i = 1,2$, is equipped with a linear array of length $2L_{R_i}$. The lengths are normalized by the wavelength of the carrier, and thus are unitless. For each array, define a local coordinate system with origin at the midpoint of the array and z-axis aligned along the length of the array. Let $\theta_{T_j} \in [0,\pi)$ denote the elevation angle relative to the $T_j$ array, and $\theta_{R_i}$ denote the elevation angle relative to the $R_i$ array. It can be shown that the field pattern radiated from the $T_j$ array will depend on $\theta_{T_j}$ only through $cos \theta_{T_j}$, thus let $t_j=cos \theta_{T_j} \in (-1,1]$, and likewise $\tau_i=cos\theta_{R_i} \in (-1,1]$. Denote the current distribution on the $T_j$ array as $x_j(p_j)$, where $p_j \in [-L_{T_j},L_{T_j}]$ is the position along the lengths of the array, and $x_j: [-L_{T_j},L_{T_j}]\rightarrow \mathbb{C}$ gives the magnitude and phase of the current. The current distribution, $x_j(p_j)$, is the transmit signal controlled by $T_j$, which we constrain to be square integrable. Likewise we denote the received current distribution on the $R_i$ array as $y_i(q_i),q_i \in [-L_{R_i},L_{R_i}]$.

By analyzing the current signal and the channel response kernel and working with the array and scattering responses, we can express the channel model by the array and scattering responses in the array domain, that is the transmit and receive signals are expressed as the current distributions excited along the array.
We can leverage the transmit and receive array responses to transform from the array domain to the wave-vector domain.

Let $T_j$ be the space of all field distributions that transmitter $T_j$'s array of length $L_{T_j}$ can radiate towards the available scattering clusters, $\Psi_{T_{jj}} \cup \Psi_{T_{ij}}$. Note that $T_j$ is the space of field distributions array-limited to $L_{T_j}$ and wavevector-limited to $\Psi_{T_{jj}} \cup \Psi_{T_{ij}}$. Likewise let $R_i$ be the space of field distributions that can be incident on receiver $R_i$ from the available scattering clusters, $\Psi_{R_{ii}} \cup \Psi_{R_{ij}}$, and resolved by an array of length $L_{R_i}$. The dimension of these array-limited and wavevector-limited transmit and receive spaces are, respectively, 
\begin{equation}
dim T_j = 2L_{T_j}|\Psi_{T_{jj}} \cup \Psi_{T_{ij}}|,
\end{equation}
\begin{equation}
dimR_i=2L_{R_i}|\Psi_{R_{ii}} \cup \Psi_{R_{ij}}|.
\end{equation}
Define the operator $H_{ij}:T_j \to R_i$ by
\begin{equation}
(H_{ij}X_j)(\tau)=\int_{\Psi_{T_{ij}} \cup \Psi{T_{jj}}} H_{ij}(\tau,t)X_j(t)dt,\tau \in \Psi_{R_{ij}} \cup \Psi_{R_{ii}}.
\end{equation}
We can now rewrite the channel model in the array domain as a model in the wave-vector domain as 
\begin{equation}
Y_1 = H_{11}X_1+H_{12}X_2+Z_1,
\end{equation}
\begin{equation}
Y_2=H_{22}X_2+H_{21}X_1+Z_2,
\end{equation}
where $X_j \in T_j$ for $j = 1,2$ is the wavevector signal transmitted by $T_j$, and $Y_i,Z_i \in R_i$ for $i=1,2$ is the wavevector signal received by $R_i$, and $Z_i $ is the additive noise. 

Let $R(H_{ij}) \in R_i$ denote the range of scattering operator $H_{ij}$, $R(H_{ij})^\perp \in R_i$ denote the orthogonal complement of $R(H_{ij})$. Let $N(H_{ij}) \in T_{j}$ denote the nullspace of $H_{ij}$, and $N(H_{ij})^\perp$ denote its orthogonal complement. We can thus obtain the following result:
\begin{equation}
\dim R(H_{ij}) = \dim N(H_{ij})^\perp = \min\{2L_{T_j}|\Psi_{T_{ij}}|,2L_{R_i}|\Psi_{R_{ij}}|\},
\end{equation}
\begin{equation}
\dim N(H_{12}) = 2L_{T_2}|\Psi_{T_{22}} \setminus \Psi_{T_{12}}|+2(L_{T_2}|\Psi_{T_{12}}|-L_{R_1}|\Psi_{R_{12}}|)^+,
\end{equation}
\begin{equation}
\dim N(H_{21}) = 2L_{T_1}|\Psi_{T_{11}} \setminus \Psi_{T_{21}}|+2(L_{T_1}|\Psi_{T_{21}}|-L_{R_2}|\Psi_{R_{21}}|)^+,
\end{equation}
\begin{equation}
\dim R(H_{11})^\perp = 2L_{R_1}|\Psi_{R_{12}} \setminus \Psi_{R_{11}}|+2(L_{R_1}|\Psi_{R_{11}}|-L_{T_1}|\Psi_{T_{11}}|)^+,
\end{equation}
\begin{equation}
\dim R(H_{22})^\perp = 2L_{R_2}|\Psi_{R_{21}} \setminus \Psi_{R_{22}}|+2(L_{R_2}|\Psi_{R_{22}}|-L_{T_2}|\Psi_{T_{22}}|)^+,
\end{equation}

\section{Spatial Degrees of Freedom Analysis}
With the PBT channel model defined in the previous section, we come to the following theorem:

\textit{Theorem 1}: Let $d_1$ and $d_2$ be the spatial degrees of freedom of Flow1 and Flow2 respectively. The spatial degrees-of-freedom region, $D_{FD}$, of the three-node full-duplex channel is the convex hull of all spatial degrees-of-freedom tuples, $(d_1, d_2)$, satisfying
\begin{equation}
d_1\le d_1^{\max} = 2\min \{L_{T_1}|\Psi_{T_{11}}|, L_{R_1}|\Psi_{R_{11}}|\},
\end{equation}
\begin{equation}
d_2\le d_2^{\max} = 2\min \{L_{T_2}|\Psi_{T_{22}}|, L_{R_2}|\Psi_{R_{22}}|\},
\end{equation}
\begin{equation}
\begin{multlined}
d_1+d_2\le d_{sum}^{\max}=2\min \{L_{T_2}|\Psi_{T_{22}}\setminus\Psi_{T_{12}}|+L_{R_1}|\Psi_{R_{11}}\setminus\Psi_{R_{12}}|+ \\
\max\{L_{T_2}|\Psi_{T_{12}}|, 
L_{R_1}|\Psi_{R_{12}}|\}, 
L_{T_1}|\Psi_{T_{11}}\setminus\Psi_{T_{21}}|+\\
L_{R_2}|\Psi_{R_{22}}\setminus\Psi_{R_{21}}|+
\max\{L_{T_1}|\Psi_{T_{21}}|, 
L_{R_2}|\Psi_{R_{21}}|\}\}.
\end{multlined}
\end{equation}

\begin{figure}[h]
\centering
\includegraphics[scale=0.5]{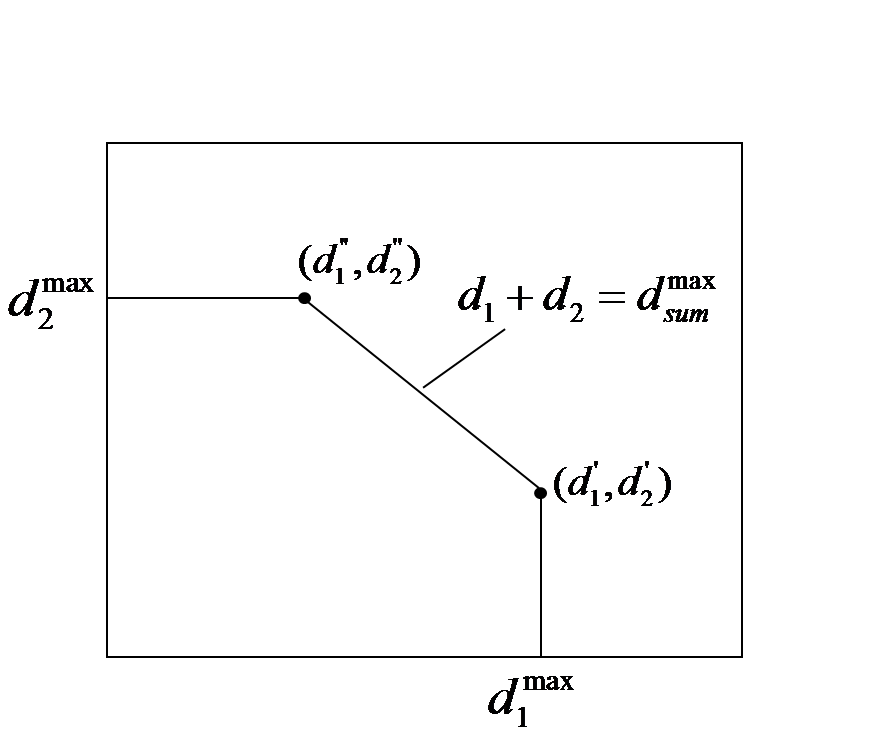}
\caption{Degree-of-freedom Region of the 3-Node Model}
\end{figure}

\subsection{Achievability}
We can establish achievability of $D_{FD}$ by the following two lemmas. The first lemma shows the achievability of two specific spatial degrees-of-freedom tuples, and the second lemma shows that these tuples are indeed the corner points of $D_{FD}$. All other points within $D_{FD}$ are thus achievable by time sharing between the schemes that achieve corner points.

\textit{Lemma 1}: The spatial degree-of-freedom tuples $(d'_1,d'_2)$ and $(d''_1,d''_2)$ are achievable, where
\begin{equation}
d'_1=\min\{2L_{T_1}|\Psi_{T_{11}}|, 2L_{R_1}|\Psi_{R_{11}}|\},
\end{equation}
\begin{equation}
\begin{split}
d'_2=\min\{d_{T_2},\delta_{R_2}\}1(L_{T_1}|\Psi_{T_{11}}|\ge L_{R_1}|\Psi_{R_{11}}|)\\ +\min\{\delta_{T_2},d_{R_2}\}1(L_{T_1}|\Psi_{T_{11}}| < L_{R_1}|\Psi_{R_{11}}|),
\end{split}
\end{equation}
\begin{equation}
\begin{split}
d''_1=\min\{d_{T_1},\delta_{R_1}\}1(L_{R_2}|\Psi_{R_{22}}| < L_{T_2}|\Psi_{T_{22}}|)\\ 
+\min\{\delta_{T_1},d_{R_1}\}1(L_{R_2}|\Psi_{R_{22}}|\ge L_{T_2}|\Psi_{T_{22}}|),
\end{split}
\end{equation}
\begin{equation}
d''_2=\min\{2L_{T_2}|\Psi_{T_{22}}|, 2L_{R_2}|\Psi_{R_{22}}|\},
\end{equation}
where $d_{T_1}$, $d_{T_2}$, $d_{R_1}$, $d_{R_2}$, $\delta_{T_1}$, $\delta_{T_2}$, $\delta_{R_1}$, and $\delta_{R_2}$ are defined in (20-27) at the bottom of this page. 1(arg) is an indicator function that evaluates to one if the argument is true, and otherwise evaluates to zero.

\textit{Sketch of Proof}: We consider the achievability of $(d'_{1},d'_{2})$, since the case of $(d''_{1},d'_{2})$ is analogous.
The full proof is omitted due to lack of space, and we show the achievability of $(d'_{1},d'_{2})$ under the conditions of $L_{T_1}|\Psi_{T_{11}}| \ge L_{R_1}|\Psi_{R_{11}}|$, $d_{T_2} \le \delta_{R_2}$, $L_{T_2}|\Psi_{T_{22}}\cap\Psi_{T_{12}}| \ge (L_{T_2}|\Psi_{T_{12}}|-L_{R_1}|\Psi_{R_{12}}|)^++L_{R_1}|\Psi_{R_{12}}\setminus\Psi_{R_{11}}|$ and $L_{T_2}|\Psi_{T_{12}}| \ge L_{R_1}|\Psi_{R_{12}}|$. In this situation, (14) and (15) are now
\begin{equation}
d'_{1}=2L_{R_1}|\Psi_{R_{11}}|,
\end{equation}
\begin{equation}
\begin{split}
d'_{2}= 2L_{T_2}|\Psi_{T_{22}}\setminus\Psi_{T_{12}}|+2L_{T_2}|\Psi_{T_{12}}|-2L_{R_1}|\Psi_{R_{12}}|+ \\ 2L_{R_1}|\Psi_{R_{12}}\setminus\Psi_{R_{11}}|,
\end{split}
\end{equation}
At $(d'_1,d'_2)$, Flow 1 has maximum degrees of freedom, thus $d'_1 = \min\{2L_{T_1}|\Psi_{T_{11}}|, 2L_{R_1}|\Psi_{R_{11}}|\} = 2L_{R_1}|\Psi_{R_{11}}|.$

The wavevector transmitted by $T_1$ and received at $R_1$, $H_{11}X_1$, is in $R(H_{11})$. If $T_2$ can construct the transmitted wavevector signal, $X_2$, such that $H_{12}X_2 \in R(H_11)^\perp$, then $H_{11}X_1 \perp H_{12}X_2$, and thus $T_2$ does not impede the recovery of the $d'_1$ signal from $T_1$ at $R_1$. Let $P_{12} = H_{12}^{\leftarrow}(R(H_{11})^\perp) \subseteq T_2$ denote the preimage of $R(H_{11})^\perp$ under $H_{12}$. Then construct $X_2$ such that $X_2 \in P_{12}$ guarantees $H_{11}X_1 \perp H_{12}X_2$. The specified conditions imply that $R(H_{11})^\perp \subseteq R(H_{12})$, thus $\dim P_{12}=\dim N(H_{12})+\dim R(H_{11})^\perp = 2L_{T_2}|\Psi_{T_{22}}\setminus\Psi_{T_{12}}|+2L_{T_2}|\Psi_{T_{12}}|-2L_{R_1}|\Psi_{R_{12}}|+ 2L_{R_1}|\Psi_{R_{12}}\setminus\Psi_{R_{11}}|=d'_2$.

Thus $T_2$ can transmit the required $d'_2$ symbols along each basis function of any orthonormal basis of $P_{12}$ without interering $R_1$. 

On the other hand, $d'_2 \le \min\{2L_{T_2}|\Psi_{T_{22}}|,2L_{R_2}|\Psi_{R_{22}}|\},$ thus $R_2$ can recover the $d'_2$ of the symbols transmitted from $T_2$. Thus the proof of this case is complete.

\textit{Lemma 2}: The degree-of-freedom pairs $(d'_1,d'_2)$ and $(d''_1,d''_2)$ are the corner points of $D_{FD}$, which are
\begin{equation}
(d'_1,d'_2)=(d^{\max}_1, \min\{d^{\max}_2,d^{\max}_{sum}-d^{\max}_1\}),
\end{equation}
\begin{equation}
(d''_1,d''_2)=(\min\{d^{\max}_1, d^{\max}_{sum}-d^{\max}_2\},d^{\max}_2).
\end{equation}

\textit{Sketch of Proof}: By checking the left and right sides of (20) and (21) under all possible conditions, we would observe equality in all cases. The procedure of computation is omitted.

\begin{strip}
\begin{equation}
d_{T_2}=2L_{T_2}|\Psi_{T_{22}}\setminus\Psi_{T_{12}}|+2\min\{L_{T_2}|\Psi_{T_{22}}\cap\Psi_{T_{12}}|,(L_{T_2}|\Psi_{T_{12}}|-L_{R_1}|\Psi_{R_{12}}|)^++L_{R_1}|\Psi_{R_{12}}\setminus\Psi_{R_{11}}|\},
\end{equation}
\begin{equation}
d_{T_1}=2L_{T_1}|\Psi_{T_{11}}\setminus\Psi_{T_{21}}|+2\min\{L_{T_1}|\Psi_{T_{11}}\cap\Psi_{T_{21}}|,(L_{T_1}|\Psi_{T_{21}}|-L_{R_2}|\Psi_{R_{21}}|)^++L_{R_2}|\Psi_{R_{21}}\setminus\Psi_{R_{22}}|\},
\end{equation}
\begin{equation}
d_{R_1}=2L_{R_1}|\Psi_{R_{11}}\setminus\Psi_{R_{12}}|+2\min\{L_{R_1}|\Psi_{R_{11}}\cap\Psi_{R_{12}}|,
(L_{R_1}|\Psi_{R_{12}}|-L_{T_2}|\Psi_{T_{12}}|)^++L_{T_2}|\Psi_{T_{12}}\setminus\Psi_{T_{22}}|\},
\end{equation}
\begin{equation}
d_{R_2}=2L_{R_2}|\Psi_{R_{22}}\setminus\Psi_{R_{21}}|+2\min\{L_{R_2}|\Psi_{R_{22}}\cap\Psi_{R_{21}}|,
(L_{R_2}|\Psi_{R_{21}}|-L_{T_1}|\Psi_{T_{21}}|)^++L_{T_1}|\Psi_{T_{21}}\setminus\Psi_{T_{11}}|\},
\end{equation}
\begin{equation}
\delta_{T_2}=2L_{T_2}|\Psi_{T_{22}}\setminus\Psi_{T_{12}}|+2\min\{L_{T_2}|\Psi_{T_{22}}\cap\Psi_{T_{12}}|,L_{T_2}|\Psi_{T_{12}}|-[L_{T_1}|\Psi_{T_{11}}|-(L_{R_1}|\Psi_{R_{11}}\setminus\Psi_{R_{12}}|+(L_{R_1}|\Psi_{R_{12}}|-L_{T_2}|\Psi_{T_{12}}|)^+)]\},
\end{equation}
\begin{equation}
\delta_{T_1}=2L_{T_1}|\Psi_{T_{11}}\setminus\Psi_{T_{21}}|+2\min\{L_{T_1}|\Psi_{T_{11}}\cap\Psi_{T_{21}}|,L_{T_1}|\Psi_{T_{21}}|-[L_{T_2}|\Psi_{T_{22}}|-(L_{R_2}|\Psi_{R_{22}}\setminus\Psi_{R_{21}}|+(L_{R_2}|\Psi_{R_{21}}|-L_{T_1}|\Psi_{T_{21}}|)^+)]\},
\end{equation}
\begin{equation}
\delta_{R_1}=2L_{R_1}|\Psi_{R_{11}}\setminus\Psi_{R_{12}}|+2\min\{L_{R_1}|\Psi_{R_{11}}\cap\Psi_{R_{12}}|,L_{R_1}|\Psi_{R_{12}}|
-[L_{R_2}|\Psi_{R_{22}}|-(L_{T_2}|\Psi_{T_{22}}\setminus\Psi_{T_{12}}|+(L_{T_2}|\Psi_{T_{12}}|-L_{R_1}|\Psi_{R_{12}}|)^+)]\},
\end{equation}
\begin{equation}
\delta_{R_2}=2L_{R_2}|\Psi_{R_{22}}\setminus\Psi_{R_{21}}|+2\min\{L_{R_2}|\Psi_{R_{22}}\cap\Psi_{R_{21}}|,L_{R_2}|\Psi_{R_{21}}|
-[L_{R_1}|\Psi_{R_{11}}|-(L_{T_1}|\Psi_{T_{11}}\setminus\Psi_{T_{21}}|+(L_{T_1}|\Psi_{T_{21}}|-L_{R_2}|\Psi_{R_{21}}|)^+)]\}.
\end{equation}
\end{strip}

\subsection{Converse}
To establish the converse part of Theorem 1, we must show that the region $D_{FD}$, which we have shown is achievable, is also an outer bound on the degrees-of-freedom. This is equivalent to establish an outer bound on the sum degrees-of-freedom, which coincides with $d^{\max}_{sum}$. Thus lemma 4 demonstrates that the converse is true.

\textit{Lemma 4}: $d_1+d_2\le d_{sum}^{\max}=
2\min \{L_{T_2}|\Psi_{T_{22}}\setminus\Psi_{T_{12}}|+L_{R_1}|\Psi_{R_{11}}\setminus\Psi_{R_{12}}|+ 
\max\{L_{T_2}|\Psi_{T_{12}}|, 
L_{R_1}|\Psi_{R_{12}}|\}, 
L_{T_1}|\Psi_{T_{11}}\setminus\Psi_{T_{21}}|+
L_{R_2}|\Psi_{R_{22}}\setminus\Psi_{R_{21}}|+
\max\{L_{T_1}|\Psi_{T_{21}}|, 
L_{R_2}|\Psi_{R_{21}}|\}\}.$

\textit{Sketch of Proof}: We can employ a genie such that the base station scattering intervals are expanded and the arrays are lengthened, and denote the maximum of the $T_2$ and $R_1$ signaling dimensions as $dim_1$. In the analogous manner, we can employ the genie on the user side and construct $dim_2$ as the maximum of $T_1$ and $R_2$ signaling dimensions. Lemma 4 thus follows from the proof in ~\cite{c4} and ~\cite{c6}.

\section{Special Cases}
We have characterized $D_{FD}$, the degrees-of-freedom region achievable by a full-duplex base-station which uses spatial isolation to avoid self-interference while transmitting uplink signal and simultaneously receiving. Now we compare the full-duplex scenarios with the half-duplex counterparts. We also compare this full-duplex scenario with self- and inter-node interference with the scenario with only self interference. We denote the degree-of-freedom region for the latter scenario as $D_{FD'}$.

We first provide the half-duplex degrees-of-freedom region, $D_{HD}$. The half-duplex achievable region is characterized by
\begin{equation}
d_1 \le \alpha \min\{{2L_{T_1}}|\Psi_{T_{11}}|,2L_{R_1}|\Psi_{R_{11}}|\},
\end{equation}
\begin{equation}
d_2 \le (1-\alpha) \min\{{2L_{T_2}}|\Psi_{T_{22}}|,2L_{R_2}|\Psi_{R_{22}}|\},
\end{equation}
where $\alpha \in [0,1]$ is the time sharing parameter.

As shown in previous results~\cite{c2}, the degrees-of-freedom region of the full-duplex case with only self-interference, $D_{FD'}$, is characterized by
\begin{equation}
d_1\le d_1^{\max} = 2\min \{L_{T_1}|\Psi_{T_{11}}|, L_{R_1}|\Psi_{R_{11}}|\},
\end{equation}
\begin{equation}
d_2\le d_2^{\max} = 2\min \{L_{T_2}|\Psi_{T_{22}}|, L_{R_2}|\Psi_{R_{22}}|\},
\end{equation}
\begin{equation}
\begin{multlined}
d_1+d_2\le d_{sum}^{\max}=2L_{T_2}|\Psi_{T_{22}}\setminus\Psi_{T_{12}}|+2L_{R_1}|\Psi_{R_{11}}\setminus\Psi_{R_{12}}|\\ +2\max\{L_{T_2}|\Psi_{T_{12}}|, 
L_{R_1}|\Psi_{R_{12}}|\}, 
\end{multlined}
\end{equation}

It's easy to see that $D_{HD}\subseteq D_{FD} \subseteq D_{FD'}$, and we are looking for cases in which $D_{HD}\subset D_{FD}$, and $D_{FD} = D_{FD'}$. In the previous case spatial isolation performs strictly better than time or frequency division, in the latter case the inter-node interference does not pose an effect. We specifically consider the following three cases: the fully-overlapped environment, the symmetric spread environment with equivalent array sizes, and the symmetric spread environment with different array sizes.

\textit{A. Fully-Overlapped Environment}

In this case, the self-interference backscattering intervals completely overlap the forward scattering intervals of the signals-of-interest. The directions of departure from each of the transmitters that scatter to the intended downlink receiver are identical to the directions of departure that backscatter to the corresponding receivers as self-interference, so that $\Psi_{T_{11}}=\Psi_{T_{12}}$ and $\Psi_{T_{21}}=\Psi_{T_{22}}$. We assume each of the scattering intervals are of size $|\Psi|$, so that $|\Psi_{T_{11}}|=|\Psi_{T_{12}}|=|\Psi_{R_{12}}|=|\Psi_{R_{22}}|=|\Psi_{R_{11}}|=|\Psi_{T_{22}}|=|\Psi_{T_{21}}|=|\Psi_{R_{21}}|=|\Psi|.$ We also assume that the base station arrays have equal length $2L_{R_1}=2L_{T_2}=2L_{BS}$, and the user arrays have equal length $2L_{T_1}=2L_{R_2}=2L_{USR}$. Hence it can be evaluated that the full-duplex degrees-of-freedom region with node interference, $D_{FD}$, is
\begin{equation}
\begin{split}
d_i \le |\Psi| min\{2L_{BS},2L_{USR}\}, i = 1,2;\\ 
d_1+d_2 \le |\Psi|\min\{2L_{BS},2L_{USR}\};
\end{split}
\end{equation}
the full-duplex degrees-of-freedom region without node-interference, $D_{FD'}$, is simply 
\begin{equation}
d_i \le |\Psi| min\{2L_{BS},2L_{USR}\}, i = 1,2; d_1+d_2 \le 2L_{BS}|\Psi|;
\end{equation}
and the half-duplex degrees-of-freedom region, $D_{HD}$, is now 
\begin{equation}
d_1+d_2 \le |\Psi|\min\{2L_{BS},2L_{USR}\}.
\end{equation}

It follows from (35-37) that in this case, $D_{HD} = D_{FD} \subset D_{FD'}$ when $2L_{BS} > 2L_{USR}$, else $D_{HD}=D_{FD}=D_{FD'}$. It can be calculated that the gain of full-duplex with only self-interference is $(2L_{BS}-2L_{USR})|\Psi|$ over half-duplex, and when the base station arrays are no smaller than twice as long as the user arrays, the degrees-of-freedom is larger than the triangular region of degrees-of-freedom of a half-duplex model. 

For the half-duplex and full-duplex with inter-node interference cases, the degrees-of-freedom increases only when both base station arrays and user arrays increase. Thus these two models are equivalent in this situation.

\textit{B. Symmetric Spread with Equal Array Lengths}

In this case, the self-interference scattering and the forward-scattering are not fully overlapped, so that the impact of the overlap of scattering intervals on full-duplex performance can be emphasized. To reduce the number of variables, suppose that the base station arrays and user arrays all have the same sizes of length $2L$, that is $2L_{T_1}=2L_{R_1}=2L_{T_2}=2L_{R_2}=2L.$ We also assume that the size of the forward scattering intervals to/from the intended receiver/transmitter is the same for all arrays $|\Psi_{T_{11}}|=|\Psi_{T_{22}}|=|\Psi_{R_{11}}|=|\Psi_{R_{22}}|=|\Psi_{fwd}|,$ and denote this interval as the forward interval. Similarly, the size of backscattering intervals is the same for all arrays in the same manner $|\Psi_{T_{12}}|=|\Psi_{T_{21}}|=|\Psi_{R_{12}}|=|\Psi_{R_{21}}|=|\Psi_{back}|,$ and denote this interval as the backscatter interval. The amount of overlap between the backscattering and forward scattering is the same at the intended transmitters as the corresponding receivers, so that $|\Psi_{T_{22}} \cap \Psi_{T_{12}}|=|\Psi_{R_{11}} \cap \Psi_{R_{12}}|=|\Psi_{T_{11}} \cap \Psi_{T_{21}}|=|\Psi_{R_{22}} \cap \Psi_{T_{21}}|=|\Psi_{fwd} \cap \Psi_{back}|=|\Psi_{fwd}|-|\Psi_{fwd}\setminus \Psi_{back}|.$

Under these conditions, the full-duplex both with and without node-interference degree-of-freedom regions, $D_{FD}$ and $D_{FD'}$, are 
\begin{equation}
d_i \le 2L|\Psi_{fwd}|,i=1,2;
\end{equation}
\begin{equation}
d_1+d_2 \le 2L(2|\Psi_{fwd} \setminus \Psi_{back}|+|\Psi_{back}|).
\end{equation}
The half-duplex region, $D_{HD}$, is
\begin{equation}
d_1+d_2 \le 2L|\Psi_{fwd}|.
\end{equation}
It follows from these two statements that $D_{HD}=D_{FD}=D_{FD'}$ if and only if $\Psi_{fwd}=\Psi_{back}$, else $D_{HD} \subset D_{FD} = D_{FD'}$. 

The full-duplex degree-of-freedom regions both with and without node-interference is rectangular if and only if $|\Psi_{back}\setminus \Psi_{fwd}| \ge |\Psi_{fwd} \cap \Psi_{back}|.$ Notice that  $2|\Psi_{back} \setminus \Psi_{forward}|$ is the dimensions useless for spatial multiplexing because $\Psi_{back} \setminus \Psi_{fwd}$ are the directions in which the base station and the user each couples to itself and not to the other. Since $|\Psi_{fwd} \cap \Psi_{back}|$ is the maximum dimension free for zero forcing the self-interference, when the condition is met, we can zero force any self- and inter-node interference without using any resource for spatial multiplexing to intended targets.

Consider a numerical example, where we have $|\Psi_{fwd}|=|\Psi_{back}|=1$, so that the overlap between them is $|\Psi_{fwd} \cap \Psi_{back}| \in [0,1]$. The following plot shows the change of degree-of-freedom region of full-duplex and half-duplex performance as the overlap changes. Notice that when $|\Psi_{fwd}|=|\Psi_{back}|$ so that $|\Psi_{fwd} \cap \Psi_{back}|=1$, $D_{FD'}$, $D_{FD}$ and $D_{HD}$ overlap, and are the triangular region. When the overlap is $|\Psi_{fwd} \cap \Psi_{back}| \in (0.5,1)$, $D_{FD'}$ and $D_{FD}$ increases towards the rectangular boundary. As $|\Psi_{fwd} \cap \Psi_{back}| \in (0,0.5]$, the full-duplex model both with and without inter-node interference is performing ideally and attains the rectangular region.
\begin{figure}[h]
\centering
\includegraphics[scale=0.3]{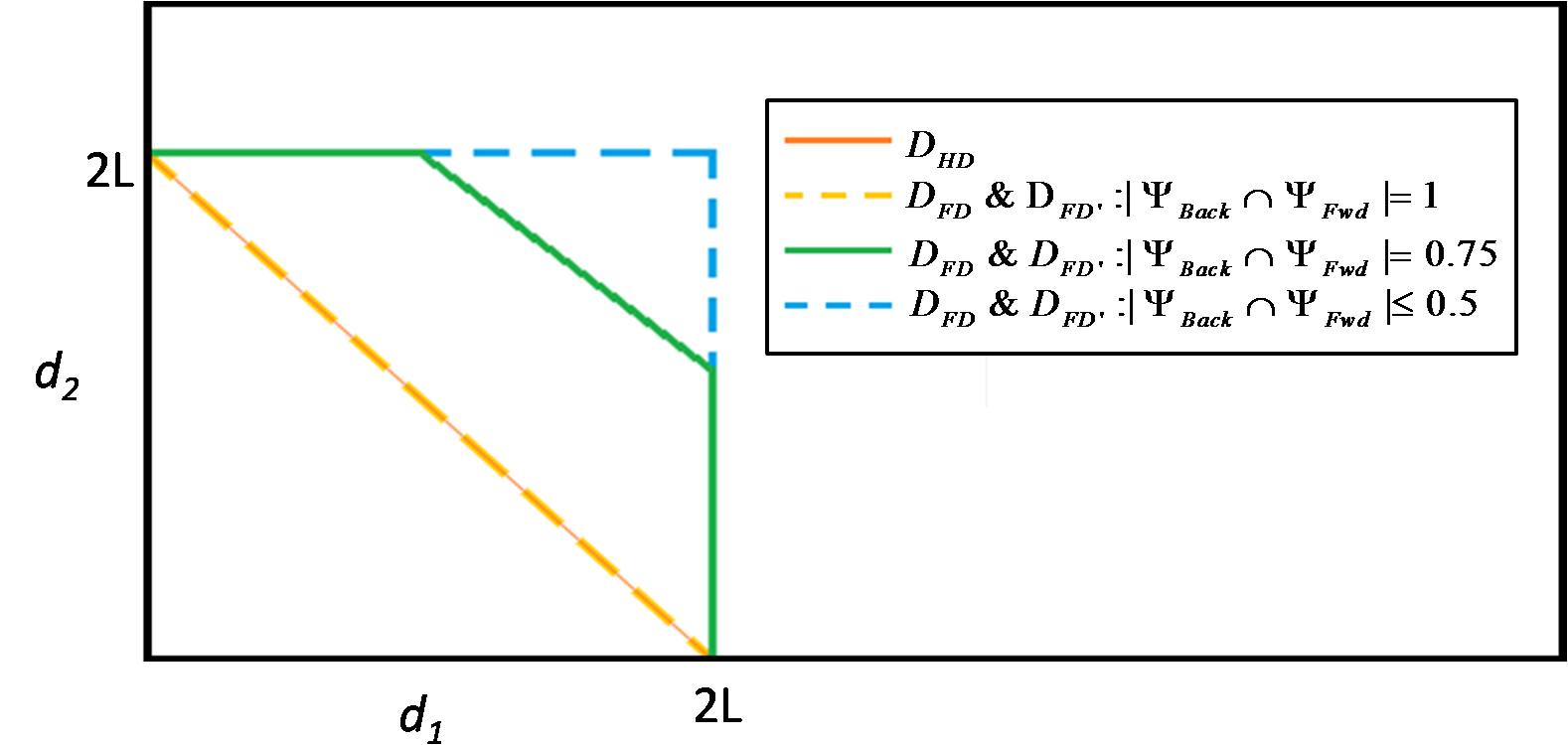}
\caption{Numerical Example of Symmetric Spread with Equal Array Lengths}
\end{figure}

\textit{C. Symmetric Spread with Different Array Lengths}

In this case, the scattering intervals are symmetric as in the previous case, so that we still have $|\Psi_{fwd}|$ and $|\Psi_{back}|$ as specified above. Now suppose the base stations arrays have equal length, $2L_{R_1}=2L_{T_2}=2L_{BS}$, and the user arrays have equal length $2L_{T_1}=2L_{R_2}=2L_{USR}$. In the previous section, we dealt with the specific case of $L_{BS}=L_{USR}$, and we would like to see if changing the array sizes will enhance the full-duplex performance. Without loss of generality, assume that $L_{USR}$ holds constant, and change $L_{BS}$ to observe its effect.

From the last case, observe that the worst case is when $|\Psi_{fwd} \setminus \Psi_{back}|=0$, so that the forward scattering and backward scattering completely overlaps, and thus the full-duplex with and without inter-node interference and half-duplex perform equivalently. In this case, we can use the conclusion from Case A, so that when $L_{BS} \ge 2L_{USR}$, the degrees-of-freedom of full-duplex with only self-interference stays rectangular. However, the degrees-of-freedom of full-duplex with inter-node interference is equal to that of the half duplex case in this case, which are both restricted by the minimum array size.

As the overlap between the forward scattering and backward scattering decreases, the conclusion of Case B demonstrates that the degree-of-freedom region of full-duplex increases from the triangle to the rectangle. 

For the full-duplex scenario with self- and inter-node interference, the degrees-of-freedom region is determined by the minimum of array sizes of base station arrays and user arrays, thus simply increasing base station arrays cannot maximize the degrees-of-freedom region. When both user arrays and base station arrays are expanded, the region of degrees-of-freedom increases and we may get the desired rectangular degrees-of-freedom region.

\section{Conclusion}
We have analyzed the spatial degrees of freedom of a full duplex model with both self-interference and inter-node interference. The outer bound that we achieved demonstrates that the degrees of freedom depends on the array sizes of base station antennas and user antennas. By comparing the full duplex models, we note that the latter model achieves better result when the base station arrays are larger. By comparing the full duplex model with both self- and inter-node interference with the half duplex model, we note that the first model performs better than the latter when we do not experience complete overlap between the forward scattering and interference scattering.

\addtolength{\textheight}{-12cm}   % This command serves to balance the column lengths

%%%%%%%%%%%%%%%%%%%%%%%%%%%%%%%%%%%%%%%%%%%%%%%%%%%%%%%%%%%%%%%%%%%%%%%%%%%%%%%%

%%%%%%%%%%%%%%%%%%%%%%%%%%%%%%%%%%%%%%%%%%%%%%%%%%%%%%%%%%%%%%%%%%%%%%%%%%%%%%%%

\end{document}